\documentstyle [11pt,epsf]{article}

\begin{document}
%
%
\newcommand{\Abs}[1]{|#1|}
\newcommand{\EqRef}[1]{(\ref{eqn:#1})}
\newcommand{\FigRef}[1]{fig.~\ref{fig:#1}}
\newcommand{\Abstract}[1]{\small
   \begin{quote}
      \noindent
      {\bf Abstract - }{#1}
   \end{quote}
    }
\newcommand{\FigCap}[2]{
\ \\
   \noindent
   Figure~#1:~#2
\\
   }
%
%
%
%
\title{Periodic orbit asymptotics for intermittent
Hamiltonian systems}
\author{Per Dahlqvist\\
Mechanics Department \\
Royal Institue of Technology, S-100 44 Stockholm, Sweden\\[0.5cm]
}
\date{}
\maketitle
\footnotetext[1]{Proceedings of the Los Alamos Center
for Nonlinear Science {\em Quantum Complexity in Mesoscopic Systems}, May 1994,
to appear in {\em Physica} {\bf D}.}

\ \\ \ \\

%
\Abstract{We address the problems in applying cycle expansions
to
bound chaotic systems, caused by e.g. intermittency
and incompleteness of the
symbolic dynamics.
We discuss zeta functions associated with weighted
evolution operators and in particular a
one-parameter family of weights relevant for
the calculation of classical resonance spectra,
semiclassical spectra and topological entropy.
For bound intermittent system we discuss an approximation of
the zeta function in terms of probabilities rather than cycle instabilities.
This approximation provides a generalization of the
fundamental part of a cycle expansion for a finite
subshift symbolic dynamics. This approach
is particularly suitable for determining
asymptotic properties of periodic orbits which are essential
for scrutinizing the semiclassical limit of Gutzwiller's
semiclassical trace formula. The Sinai billiard is used as model system.
In particular we develope a crude approximation of the semiclassical zeta
function which turns out to possess non analytical features.
We also
discuss the contribution to the semiclassical level density from the neutral
orbits.
Finally we discuss implications of our findings for the spectral form factor
and compute the asymptotic behaviour of the spectral rigidity. The result is
found to be consistent with exact quantum mechanical calculations.}

\ \\

\section{Introduction - Dilemmas in Quantum Chaology}

There is no definition of
{\em Quantum Chaology} or
{\em Quantum Chaos} which is generally agreed upon. One may
focuse on the correspondance principle and state the question
{\em How is chaos revealed in a quantum system when the Planck's constant
$\hbar \rightarrow 0$, if its classical counterpart is chaotic}.
The answer turns out to as complicated as the question is interesting.
It seems as if one cannot understand properly the transition from
quantum to classical mechanics for a system with few degrees of freedom
without considering its coupling to the environment.
This coupling may induce decoherence,
enabling quasi classical behaviour \cite{Zurek}.

One may also (and we will) focuse on the more modest question:
{\em What is the behaviour of a quantum system if its classical counterpart
is chaotic?} We thus avoid any discussion of decoherence phenomena.
Restricting our attention to bounded autonomous systems we may
choose e.g. to study quantum spectra.
There has been a lot of numerical and experimental studies of the statistical
properties of quantum spectra of classically chaotic systems.
There are some widely known conjectures
inspired by random matrix theories \cite{Boh}.
Although some of the numeric evidence is rather striking there is very
little theoretical understanding. Such an attempt naturally starts from the
semiclassical trace formula.
This formula relates the density of
eigen-energies to dynamical invariants of the
periodic orbits of a system \cite{Gut}:
\begin{equation}
      g(E)=g_{o}(E)
           +\frac{1}{i{\hbar}}\sum_{p}T_{p}\sum_{n=1}^{\infty}
            \frac{1}{\Abs{M_{p}^{n}-I}^{\frac{1}{2}}}
            \,e^{in
               \left[S_{p}/\hbar-\mu_{p}\frac{\pi}{2}\right]}
            \ \ .
                                                   \label{eqn:Gut}
\end{equation}
Here $g(E)=\mbox{Tr}\,G(q,q';E)$ is the trace of the semiclassical
Green's function, and the density of eigenstates is given by
$d(E)=-\frac{1}{\pi}\lim_{\epsilon \rightarrow 0}
\,\mbox{Im}\,g(E+i\epsilon )$.
The index $p$ labels the primitive periodic
orbits, $S_{p}$ is the action integral along the orbit, $T_{p}$
its period,
$M_{p}$ is the linearized Poincar\'{e} map around the orbit (monodromy
matrix), $\mu_{p}$ is the Maslov index, and
$g_{o}(E)$ provides the mean level distribution.
We will exclusively consider billiards and put $\hbar =m=1$. The
semiclassical limit $\hbar \rightarrow 0$ is thus replaced by
$E \rightarrow
\infty$. The action is given by $S_p=l_p \cdot \kappa(E)$ where
$\kappa$ is the momentum
$\kappa =\sqrt{2E}$, and $l_p$ is the length of the prime orbit.

The trace formula is derived through
a stationary phase approximation, except for
$g_0(E)$  which is calculated from the short time behaviour of the
semiclassical propagator. In principle $g_0(E)$ is given by an asymptotic
expansion whose leading term is the Weyl term. The division into a mean
contribution and an oscillating may turn out to be more involved than
anticipated. We will indeed encounter such problems in this paper.

It is often more convenient to study
the Gutzwiller-Voros
zeta function \cite{Vor} which for systems with two
degrees of freedom reads
\begin{equation}
     Z_{GV}(E)=\prod_{p}\prod_{m=0}^{\infty}
          \left(1-\frac{e^{il_{p}\kappa (E)-\mu_{p}\frac{\pi}{2}}}
    {\Abs{\Lambda_{p}}^{1/2} \Lambda_{p}^{m}}\right)
          \ \ .                       \label{eqn:ZGV}
\end{equation}
where $\Lambda_p$ is the expanding eigevalue of $M_p$.
$Z_{GV}$ is related to the level density by
\begin{equation}
     g(E)-g_{0}(E)=\frac{d}{dE}\ln Z(E) \ \ ,
\end{equation}
so the zeros of the zeta function are the poles of the of eq.
\EqRef{Gut}, i.e. the semiclassical eigenvalues.

It is not known whether the trace formula manages to
give the right density of states
on, or close to the real energy axis.
Whatever method one uses, only a moderate number of zeros can
be computed from a reasonable number of periodic orbits.
The exponential proliferation of periodic orbits
makes every step towards the semiclassical limit very expensive.
It therefore seems impossible in practice to
study the semiclassical limit of the semiclassical trace formula by explicit
computation of the periodic orbits.

This motivates a study of asymptotic measures of the set of periodic orbits
for bound systems, which is the main problem to be addressed in this talk.

We will make extensive use of the concept of {\em zeta functions}.
The
most succesful use of zeta functions has been achieved
for so called Axiom-A systems (a very strong definition of chaos suitable
for making mathematical theorems).
Then there are theorems about entireness of the certain classes of zeta
functions.  However, bounded systems are generally not Axiom-A,
they are generally intermittent and lack a simple symbolic dynamics,
so practical
applications of the theory of zeta functions are not very successful.
Since our main motivation is to understand fluctations in the spectra of
bound systems we are presented with another dilemma.
It will however turn out that we can make use of this inconvenient
intermittent property to formulate approximate zeta functions.

\section{A menagerie of zeta functions}

This exposition of zeta functions and the thermodynamic
formalism is rather brief.
The reader may found more details in refs \cite{flows,Rosen,CR93,AAC,PDreson}
and references therein.

\subsection{Evolution operators, trace formulas and zeta functions}

Zeta functions are introduced by considering the
evolution operator ${\cal L}_w^t$.
It describes the evolution of a phase space density
$\Phi(x)$
\begin{equation}
{\cal L}_w^t  \Phi(x)=\int w(x,t)\delta (x-f^t(y))\Phi(y)dy   \ \ .
\end{equation}

The phase space point $x$ is taken by the flow to $f^t(x)$ during time
$t$. $w(x,t)$ is a weight associated with a trajectory starting at $x$
and evolved during time $t$. It is multiplicative along the flow, that
is $w(x,t_1+t_2)=w(x,t_1)w(f^{t_1}(x),t_2)$. This ensures that the
eigenvalues are on the form $\lambda_{\alpha}^t$.

As we are only studying billiards it is convenient to use
the length traversed $l$ as {\em time} variable.

We now
compute the trace of the evolution
operator, that is, the sum of its eigenvalues
$tr {\cal L}_w^l =\sum_{\alpha} e^{ik_{\alpha}l}$.
The trace may be written as a sum over the
periodic orbits in the system
\begin{eqnarray}
tr {\cal L}_w^l  =\int w(x,t) \delta (x-f^l(x))dx= \nonumber \\
\sum_p l_p \sum_{n=1}^{\infty} w_p^n \frac{\delta(l-nl_p)}
{\Abs{det(1-M_p^n)}}  \ \ ,
\label{eqn:tracedef}
\end{eqnarray}
The trace may also be written as the
Fourier transform of the logarithmic derivative
of a {\em zeta function}
\begin{equation}
tr {\cal L}_w^l = \frac{1}{2\pi i}
\int_{-\infty}^{\infty} e^{ikl}\frac{Z_w'(k)}{Z_w(k)}dk   \ \ .
\label{eqn:traceZ}
\end{equation}
The zeta function
reads
\begin{equation}
     Z_w(k)=\prod_{p}\prod_{m=0}^{\infty}
          \left(1-w_p \frac{e^{-ikl_{p}}}
    {\Abs{\Lambda_{p}} \Lambda_{p}^{m}}\right)^{m+1}
          \ \ ,                       \label{eqn:Zw}
\end{equation}
The $k_{\alpha}$'s introduced above
should be the zeros of the zeta function. For the system we are going to
consider things will be complicated and the operator will not have a discrete
spectrum.

By using different weights $w$ one can probe different properties
of the flow. We will consider the one-parameter family of weights
$\omega = \Abs{\Lambda(x,t)}^{\tau}$.
$\Lambda(x,t)$ is the expanding eigenvalue
of the Jacobian transverse to the flow.  It is only
approximately
multiplicative along the flow but it is possible to modify it slightly so as
to become exactly multiplicative \cite{Gabor}.
However,
this is a subtlety compared to the approximations we are
going to apply.

The leading zero will generally be on the imaginary axis
$k=-ih(\tau)$, where $h(\tau)$ are generalized entropies, in the thermodynamic
formalism they are usually named {\em topological pressures}, \cite{Beck}.

Putting $\tau=0$ we obtain the {\em classical} zeta function
whose zeros yields the so called {\em resonance spectrum} or
spectrum of correlation exponents (again things are complicated if the
spectrum is not discrete)
\begin{equation}
     Z(k)=\prod_{p}\prod_{m=0}^{\infty}
          \left(1- \frac{e^{-ikl_{p}}}
    {\Abs{\Lambda_{p}} \Lambda_{p}^{m}}\right)^{m+1}
          \ \ .                       \label{eqn:Zcl}
\end{equation}
The leading zero provides the escape rate. For a bound
system the escape rate equals zero: $h(0)=0$.

The {\em topological} zeta function
is obtained by considering the case $\tau=1$.
The leading zero now gives the
topological entropy $h(1)$. The asymptotic behaviour of the
trace is
\begin{equation}
tr {\cal L}_{top}^l  \approx
\sum_p l_p \sum_{n=1}^{\infty}  \delta(l-nl_p)
\rightarrow e^{h(1) t}
\end{equation}
so that the number of cycles with periods less than $l$ is
$\sim e^{h(1) l}/h(1) l$.
To obtain this exponential increase, the contour of the Fourier transform
\EqRef{traceZ} must extend below all zeros where  the Euler product \EqRef{Zw}
converges.

Let us now study the case
$\tau=1/2$.  We then get
\begin{equation}
     Z_{sc}(k)=\prod_{p}\prod_{m=0}^{\infty}
          \left(1- \frac{e^{-ikl_{p}}}
    {\sqrt{\Abs{\Lambda_{p}}} \Lambda_{p}^{m}}\right)^{m+1}
          \ \ .                       \label{eqn:Zsc}
\end{equation}
This zeta function
is called {\em the quantum Fredholm determinant}
\cite{Rosen,CR93} and is equivalent to the
Gutzwiller-Voros zeta function in the semiclassical limit
through the identification $\kappa(E)=-k$.
Its leading zero is closer to the origin than the
topological one $0<h(1/2)<h(1)$, cf. sections 2.2 and 3.3.
This zero provides a pole in the trace
formula, and the periodic orbit sum \EqRef{tracedef} as well as the trace
formula \EqRef{Gut} diverge in the half plane above it.
This is usually
referred to as a an entropy barrier.
The Euler-product representation of the
zeta function \EqRef{Zw} also diverges above this leading zero (because
the zero is not a zero of an individual factor and an infinite product is not
allowed to converge towards zero).
We have left out the Maslov indices
but it is possible to account for them in the weight as well.
The introduction of phase indices will shift the
the leading zero towards the origin as we soon will see.


\subsection{Cycle expansions of zeta functions}

 The zeta functions have been formulated as Euler
products. These products diverge wherever they are interesting (i.e. at the
nontrivial zeros). In computations, one generally expands them
into infinite series, power series
in the case of maps
and Dirichlet series in case of flows.
Power series converge up to its first singularity and are thus better for
computations. Dirichlet series are more complicated, they may have a strip of
conditional convergence and do not neccesarily have singularities on its
border(s) of convergence.

Zeta functions are entire if the system has a Poincar\'{e} map
which is analytic, hyperbolic and have a finite subshift symbolic dynamics
\cite{Rugh}. In that case the expansion may be divided into a fundamental
part, yielding the gross structure of the spectrum,
 and {\em curvature corrections} \cite{AAC}. The standard example is
the expansion of the {\em dynamical zeta function} (i.e. the $m=0$ factor
of the zeta functions above) for a system with binary symbolic dynamics:
\begin{equation}
\begin{array}{l}
 \prod_p (1-t_p)= \\
            1-t_{1}-t_{0}                   \\
            - [t_{10}-t_{1}t_{0}]            \\
           - [t_{100}-t_{10}t_{0}] - [t_{110}-t_{1}t_{10}] \\
            - [t_{1100}-t_{1}t_{100}-t_{110}t_{0}+t_{1}t_{10}t_{0}]
            - [t_{1110}-t_{1}t_{110}]
            - [t_{1000}-t_{100}t_{0}] \\
            - \cdots \cdots
      \label{eqn:curv}
\end{array}
\end{equation}
where $t_p=\exp(-il_p k)/|\Lambda_p|^{1-\tau}$.
The expansion is organized according to increasing symbol length in an
obvious way, in the case of a map (the $l_p$'s being integers) this means
according to increasing powers of $exp(-ik)$. The expansion
is dominated by the {\em fundamental} contribution,
$1-t_{1}-t_{0}$.
For the lowest powers there is a nice interpretation why the rest of the
expansion is small \cite{AAC}.
The tail of the expansion is organized in such a way that
each square bracket contains some long orbit(s) (length $n$)
minus its approximant(s) in
terms of shorter ones. The sizes of these
{\em curvature corrections} fall of exponentially with $n$.
This leads to a pole lying beyond the
leading zero.
Similar
expansion for the full zeta function has infinite radius of convergence
(under the above conditions) such that nonleading zeros may be
extracted.

{\em Example.} Let us consider a hypothetical billiard having the Baker's map
as Poincar\'{e} map. Then $|\Lambda_p|=2^{n_p}$. Let further the periods be
discretized $T_p=n_p T_0$. Only the fundamental part will remain:
\begin{equation}
Z=1-2\frac{e^{T_0 k}}{2^{1-\tau}}
\end{equation}
with leading zero $=-i\tau \log 2 /T_0$. We see that
$h(0)=0$ as it should be for a bound system and the topological
entropy is $h(1)=\log 2 /T_0$. This is of course a very unrealistic example and
the semiclassical spectrum, $\tau =1/2$, does not look sensible at all.

If the symbolic dynamics is a more general, finite subshift (meaning
that there is a {\em finite} set of {\em pruning rules}), this
scheme may still be worked out. There is a wealth of illustrative examples in
ref. \cite{AAC}.
However, if the symbolic dynamics is not a finite subshift
there will be no similar expansion. The scheme outlined here is therefore not
applicable for a generic bound system. In this case there is not even a well
defined fundamental part.
We will suggest a solution of this dilemma in the next section.

\subsection{Zeta functions in the BER approximation}

In ref. \cite{PDreson} an approximate expression for the zeta function
is given for intermittent, ergodic Hamiltonian systems. The idea
is based on a paper
by Baladi, Eckmann and Ruelle \cite{BER} so we refer to it
as the BER approximation.

In an intermittent system laminar intervals are interupted
by chaotic outbursts. Let $\Delta_i$ be the time ellapsed between two
succesive entries
into the laminar phase. The index $i$ labels the i'th
interval.
Provided the chaotic phase is {\em chaotic enough},
the lengths of the intervals $\Delta_i$
are presumed uncorrelated,
and $\Delta$ may be considered as a stochastic variable
with probability distribution $p(\Delta)$.
The zeta functions may then be expressed in terms
of the Fourier transform of $p(\Delta)$
\begin{equation}
Z(k)\approx \hat{Z}(k)
\equiv 1-\int_{0}^{\infty}e^{-ik\Delta}p(\Delta)d\Delta   \   \
\label{eqn:ZBER}
\end{equation}
We use the "$\hat{\mbox{ }}$"  symbol to mark all quantities
derived in the BER approximation.

\begin{figure}[p]
\vspace{15cm}
\caption{a) The Sinai billiard with definitions of
of the phase space variables $\phi$ and $\alpha$.
b) The unfolded system with free directions (corridors) indicated.}
\end{figure}

\section{The Sinai billiard - Classical considerations}

The Sinai billiard \cite{Sin} consists of a unit square with a
scattering disk, having radius $R$;
$0<2R \leq 1$, centered on its midpoint.
This billiard  has a fairly simple geometry, but exhibits many
features typical for bound
chaotic system; it is intermittent and lacks a simple
symbolic dynamics.

The trajectory of a particle in
the Sinai billiard consists of laminar intervals,
(bouncing between the straight sections) interrupted by scatterings
off the central disk. The Sinai billiard seems therefore suited
for the BER approximation. The variable
$\Delta$ introduced above is simply the length
of the trajectory segment between two
disk bounces.

The material of this section, except for sect. 3.3, is developed in ref
\cite{PDSin} in more detail.

\subsection{Formulation of $p(\Delta)$ and a generalization}

We use the
disk to define the surface of section. The canonical variables
are
$(R\phi,\sqrt{2E}\sin \alpha )$ with the angles
$0< \phi \leq  2\pi $ and  $-\frac{\pi}{2}<\alpha<\frac{\pi}{2}$
defined in fig. 1a.
The  map
$(\phi,\sin \alpha ) \mapsto (\phi,\sin \alpha )$ is area preserving and
has uniform invariant density.

We will now use this uniformity to formulate an expression for
$p(\Delta )$. To that end we use a
natural partition of the surface of section defined by the topology of the
problem. Suppose a laminar segments hits the vertical walls
$n_x$ times, and the horizontal walls
$n_y$ times. Equivalently one can
unfold the billiard into a regular lattice of disks.
These disks are assigned
numbers $q=(n_x,n_y)$, as in fig 1b. Only one octant need to be considered
due to the $C_{4v}$ symmetry, so we take $n_x$ and $n_y$ positive, and
$n_x \geq n_y$. The segment under study starts out from  $q=(0,0)$.
It is obvious that only coprime $q=(n_x,n_y)$ can be reached.
Now let
$\Omega_q$ be the part of phase space $(\phi,\alpha)$ for which the trajectery
hit disk $q$.
Let us make the further approximation that all trajectories
going from $(0,0)$ to $q$ have the same length $l_q$.
We then get the following approximation for $p(\Delta)$
\begin{equation}
\hat{p}(\Delta) \approx \sum_{q} a_q(0) \delta (\Delta-l_q) \ \ ,
\end{equation}
where $a_q(0)$ is given by
\begin{equation}
a_q(0)=\frac{2}{\pi} \int_{\Omega_q}d\phi \; d(sin\;\alpha )
\ \ .
\label{eqn:aqdef}
\end{equation}

This approach is not restricted to the
$\tau=0$ case. To treat the other cases we simply average a power of the local
instability $|\Lambda(\phi,\alpha)|\approx \frac{2l_q}{Rcos (\alpha)}$ over
$\Omega_q$.  This leads to the generalized probability distributions
\begin{equation}
\hat{p}_{\tau}(\Delta) \approx \sum_{q} a_q(\tau) \delta (\Delta-l_q) \ \ ,
\label{eqn:pDelta}
\end{equation}
with
\begin{equation}
a_q(\tau)=\frac{2}{\pi}\int_{\Omega_s}
\Abs{\Lambda(\phi,\alpha)}^{\tau}d\phi
\;  d(sin\; \alpha)
\label{eqn:aq}
\end{equation}
The zeta function may now be written
\begin{equation}
\hat{Z}_{\tau}(k) = 1-\int_{0}^{\infty}e^{-ik\Delta}
\hat{p}_{\tau}(\Delta)d\Delta
\approx 1-\sum_{q} a_{q}(\tau) e^{-ikl_q} \ \ ,
 \label{eqn:Zhat}
\end{equation}

\begin{figure}
\epsffile{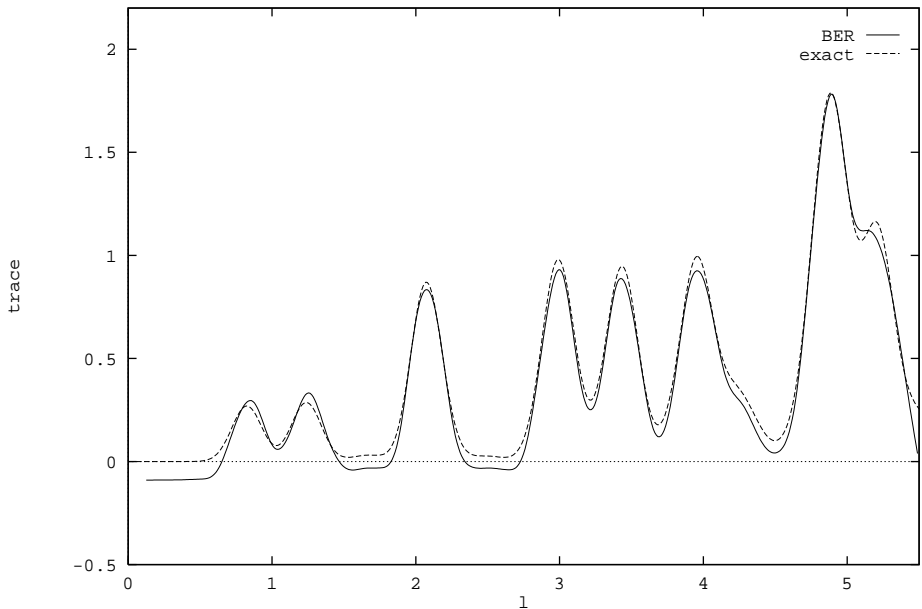}
\caption{The trace of the evolution operator (unit weight, $\tau=0$)
for the BER approximation and the explicit
periodic orbit sum. Both curves has been smoothened by a gaussian having width
$\sigma=0.1$. The disk radius is $R=0.1$. From [15].}
\end{figure}

\subsection{Comparison with periodic orbit calculations}

The formulation of $\hat{p}_{\tau}(\Delta)$ can be further refined. The region
$\Omega_q$ may divided into eight parts $\Omega_s$ with $s=(g,q)$
where $g$ depend
on the octant from which the trajectory has
arrived. This eight octants correspond to the eight elements of the symmetry
group $g \in C_{4v}$ of the billiard. The zeta function now reads
\begin{equation}
\hat{Z}_{\tau}(k)
\approx 1-\sum_{s} a_{s}(\tau) e^{-ikl_s} \ \ ,
 \label{eqn:Zhattau}
\end{equation}
This amounts to restricting the system to the fundamental domain and
study one symmetry subspace $A_1$ of the original problem, see refs.
\cite{PDSin,CEsym}.  For each $\Omega_s$ there is an orbit $\bar{s}$
periodic in the fundamental domain.
We choose
$l_s$ to be the length of this orbit.
(It may need to go through an intermediate
disk, but $l_s$ is well defined irrespective if $\bar{s}$ is pruned or not).
Other subspaces may also be considered by inserting group
characters, see \cite{PDSin,CEsym}. We will not discuss them here
to keep things as simple as possible.

At this point it is interesting to check if this approximate zeta function can
reproduce
the exact trace by means of the periodic orbit sum \EqRef{tracedef},
see fig 2.
This plot, reaching up to $l=5.5$, involves $\sim 8000$ periodic orbits.
The exponential proliferation of cycles
means that we cannot go very much higher in $l$ by explicit periodic orbit
calculations. The conclusion so far is that the BER approximation gives a very
good description, which is generally improved for larger $l$, see ref.
\cite{PDSin}. The fine scale structure can of course not be reproduced due to
the introduction of deltafunctions in section 3.1 and due to the
BER approximation itself.
The topological zeta function ($\tau=1$) was examined in ref. \cite{PDSin},
also  with good
results. It is also demonstrated in ref \cite{PDSin} that
the exact topological zeta function is obtained in the limit
$R \rightarrow 0$. The conclusion is that the method works for
any value of $\tau$.

\begin{figure}
\epsffile{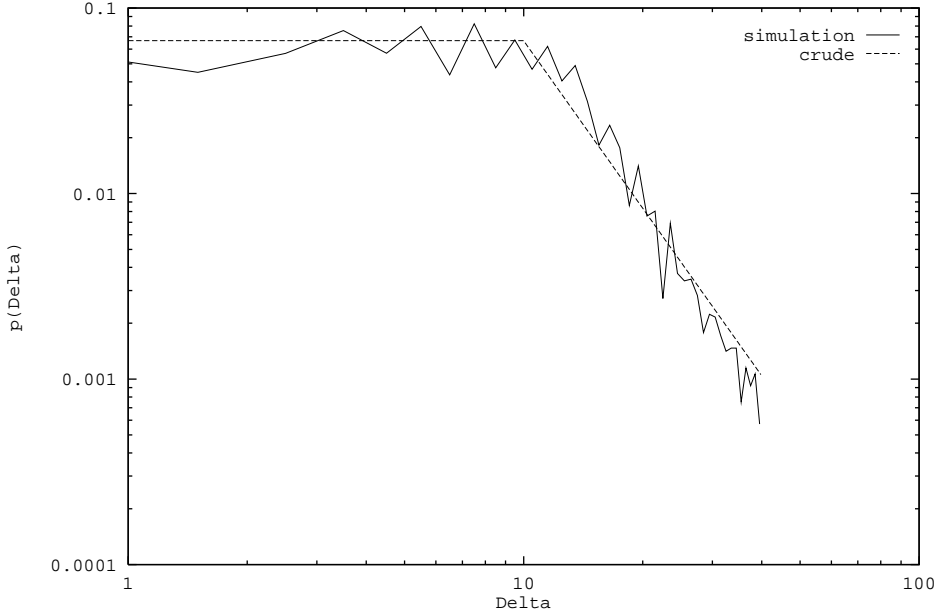}
\caption{$p(\Delta )$ from numerical simulation, compared with the crude
approximation of sect 3.3. The disk radius is $R=0.05$.}
\end{figure}

\subsection{A crude approximations to $p_{\tau}(\Delta)$}

We will now work out a very crude approximation to $p_{\tau}(\Delta)$
which will turn out to be both illustrative and useful.
For any disk radius $0<R<1/2$ there is a finite number of directions
along which a trajectory
may go without ever bouncing off a disk (see fig 1b)! We call them {\em free
directions}, or
{\em corridors}. Consider the direction vector $(x,y)$. A direction
cannot be free if $x/y$ is irrational. So we may take $(x,y)$ as
positive coprime integers $(n_x,n_y)$ such that $n_x/n_y \leq 1$.
It is a little confusing that $(n_x,n_y)$ may refer both to a
particular disk in the unfolded
system and a corridor but thus ambiguity will turn out to be rather convenient.
The direction $(n_x,n_y)$ is free
if \cite{PDSin,BSin}
\begin{equation}
2R<\frac{1}{\sqrt{n_x^2+n_y^2}}   \label{eqn:corrr} \ \ .
\end{equation}
This means that if disk $q=(n_x,n_y)$ lies inside radius $1/2R$ there are no
obstructing disks in front of it, see fig 1b.
Next we assume that the disk radius is small and set
$l_q=\sqrt{n_x^2+n_y^2}$.
The integral \EqRef{aq}  for $l_q < 1/2R$ is trivial
\begin{equation}
a_q(\tau) \approx \frac{2}{\pi} \frac{\Gamma(\frac{2-\tau}{2})^2}
{\Gamma(2-\tau)} (\frac{R}{l_q})^{1-\tau}  \   \  .
\label{eqn:aqfull}
\end{equation}
In order to find an approximate expression for $p_{\tau}(\Delta)$
we must know the
density of coprime lattice points
with respect to the radius in the first octant $d_c(r)$.
This is, to leading order, found to be $d_c(r)\approx \frac{2\pi}{13}r$, see
Appendix and fig 6.
This yields
\begin{equation}
p_{\tau}(\Delta)=\sum_q \delta (\Delta -l_q)a_q(\tau)=
d_c(\Delta) \cdot a_q(\tau) |_{l_q=\Delta}=
\frac{16}{13} \frac{\Gamma(\frac{2-\tau}{2})^2}
{\Gamma(2-\tau)} R^{1-\tau} \Delta^\tau  \   \   .
\end{equation}
Obviously the disk radius has to be small for this to apply.

If $\Delta \geq 1/2R$ the disks start the obscure each other.
The accesable disks are those along the corridors, se fig 1b.
They may be written on the form
$q_n=q'+n\cdot q''$. Here, $q''=(n_x'',n_y'')$ is a free direction and
$q'$ is one its neighbours
in the Farey sequence of order $n_x''$ \cite{Num}.
The  $a_q$'s decreases
as \cite{PDSin}
\begin{equation}
a_q (\tau) \sim \frac{1}{n^{3-3\tau /2}}  \label{eqn:aqtail}
\end{equation}
and the corresponding lengths grow linearly
\begin{equation}
l_{q_n} \approx const+n\cdot q_{free}
\end{equation}
The last two equations impliy that
\begin{equation}
p_\tau (\Delta )  \sim \frac{1}{\Delta^{3-3\tau /2}}  \;  \;  \;  \; \;  \;
\Delta \gg \frac{1}{2R}
\end{equation}
Of course, there is a transition region around $\Delta \sim 1/2R$ but if we
accept a kink at $\Delta=1/2R$ and require
$p_{\tau}(\Delta)$ to be continous we obtain
\begin{equation}
p_{\tau}(\Delta) \approx
p_{\tau}^{crude}(\Delta)= \left\{ \begin{array}{ll}
\frac{4}{3} \frac{\Gamma(\frac{2-\tau}{2})^2}
{\Gamma(2-\tau)} R^{1-\tau} \Delta^\tau  & \Delta \leq 1/2R \\
\frac{4}{3} \frac{\Gamma(\frac{2-\tau}{2})^2}
{\Gamma(2-\tau)} \frac{2^{\tau/2-3}R^{-\tau/2-2}}{\Delta^{3-3\tau /2}} &
\Delta > 1/2R  \end{array}  \right.    \label{eqn:pcrude}
\end{equation}
We have multiplied the prefactors with a factor $13/12$ so that
$Z_{\tau=0}^{crude}(0)=0$, our approximations otherwise violate the
normalization of $p(\Delta)$.
In fig. 3 we
compare this expression with a numerical simulation of the billiard for the
case $\tau =0$. Our crude approximation is a good smoothened approximation to
$p(\Delta)$.

The most obvious length scale in the Sinai billiard
is the length of the shortest periodic orbit
$l_0 \approx 1$. We have now learned that there is another
important  length scale, $l_{scale}=1/2R$, given by the kink above, being
the mean free path in the unfolded system.
We will subsequently understand that this is a very
important length scale even for the quantum problem.

A crude estimation of the zeta function $Z_{\tau}^{crude}(k)$ is obtained by
a Fourier transform of $p_{\tau}^{crude}(\Delta)$. What can we say about
its analytic structure?

\begin{figure}
\epsffile{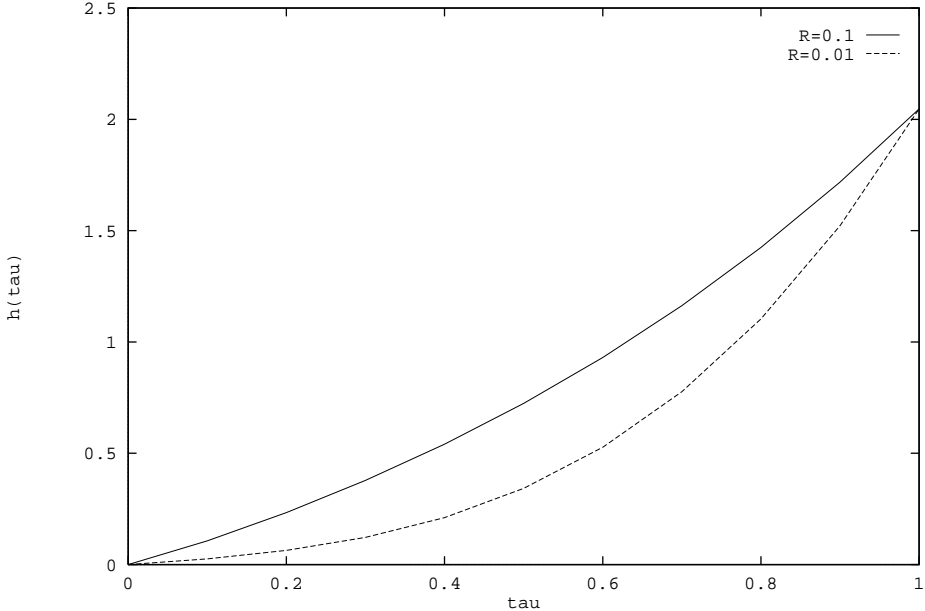}
\caption{The generalized entropies $h(\tau)$ as given by the leading zero
of the crude approximation of the zeta function.}
\end{figure}

The leading zero $k_0=-ih(\tau)$
as obtained from this zeta function is plotted versus
$\tau$ for two different disk radii in fig. 4.
We see that the  behaviour is not at all linear as
was the case in the example of section 2.2, which is the case only for
uniformly hyperbolic systems. The Sinai billiard is anything but unifomly
hyperbolic, the intermittency becomes more and more pronounced as $R$
decreases and the deviation from a linear function increases accordingly. Note
that  a linear dependence is often assumed in some
oversimplified (erroneous) estimations of the {\em entropy barrier} in
the litterature. 

For small $\tau$,that is, when the
zero is close to the origin, the zeta function is dominated by the large $l$
tails.  Further
away from the origin it is dominated by the
closest disk. The topological entropy $h(1)$
obtained this way is 5\% wrong in the limit $R \rightarrow 0$, see ref.
\cite{PDSin}.  The reason is twofold: the ad hoc renormalization of the
prefactor, and secondly, for this large $\tau$ the value is dominated by the
closest disks.

Equipped with our crude estimation of the zeta function we can insert it onto
\EqRef{traceZ} and the compare the trace obtained with the result of section
3.2. The comparison is made in fig. 5. Wee se that the
crude approximation give a good average
description.  We see that when $l>l_{scale}$ the curve starts to approach
an asymptotic behaviour. What is this limiting behaviour.

The power law tail of $p_{\tau}(\Delta)$ introduces a branch cut
of the zeta function
along
the positive imaginary axis. The asymptotics of the Fourier transform
\EqRef{traceZ} is
governed by this cut. Let us be slightly more general and
consider an arbitrary
(generalized) distribution $p_{\tau}(\Delta)$ with a power law tail
$\sim 1/\Delta^m$. Then an elementary calculation yields
\begin{equation}
\frac{1}{2\pi i}
\int_{-\infty}^{\infty} e^{ikl}\frac{Z_w'(k)}{Z_w(k)}dk \sim \left\{
\begin{array}{lll}
1-\frac{p_{\tau}(\Delta_0)\Delta_0^m}
{(m-1)\int \Delta  p_{\tau}(\Delta)d\Delta } \frac{1}{l^{m-2}} &
 Z(0)=0   &   m >2  \\
-\frac{p_{\tau}(\Delta_0)\Delta_0^m}
{\int  p_{\tau}(\Delta)d\Delta -1} \frac{1}{l^{m-1}}
&  Z(0) \neq 0 & m >1 \end{array}  \right. \label{eqn:tras}
\end{equation}
$\Delta_0$ is any point in the tail of $p_{\tau}(\Delta)$. The results are
valid also for non integer $m$.
To obtain these results we let
the contour of the Fourier integral run
along the real axis.
To get consistency with
the explicit periodic orbit sums \EqRef{tracedef} for
$\tau > 0$ one has
to extend the integration below all the zeros
which will pick up up exponentially
increasing terms.

Note that there will be a sudden change in the asymptotic behaviour when $\tau
=0$. One say that a phase transition occurs at this point; the exponential
behaviour of the trace
is taken over by a power law. When $\tau=0$ the cut reaches down to
the leading zero preventing exponential mixing.

\begin{figure}
\epsffile{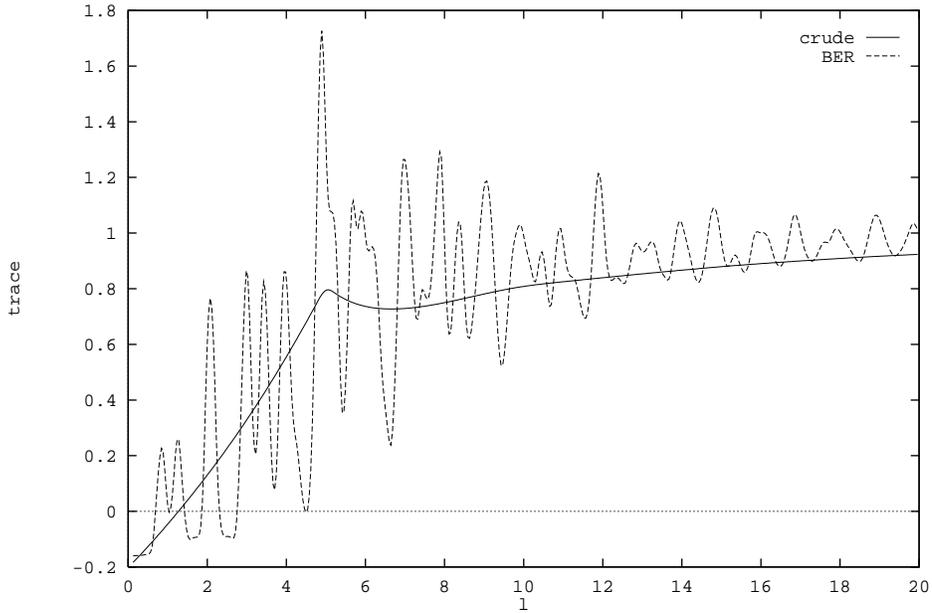}
\caption{The trace of the evolution operator ($\tau=0$),
calculated from the crude approximation (sect. 3.3)
and the more refined approximations (sect. 3.2) of the zeta functions,
for disk radius $R=0.1$.} \end{figure}

\begin{figure}
\epsffile{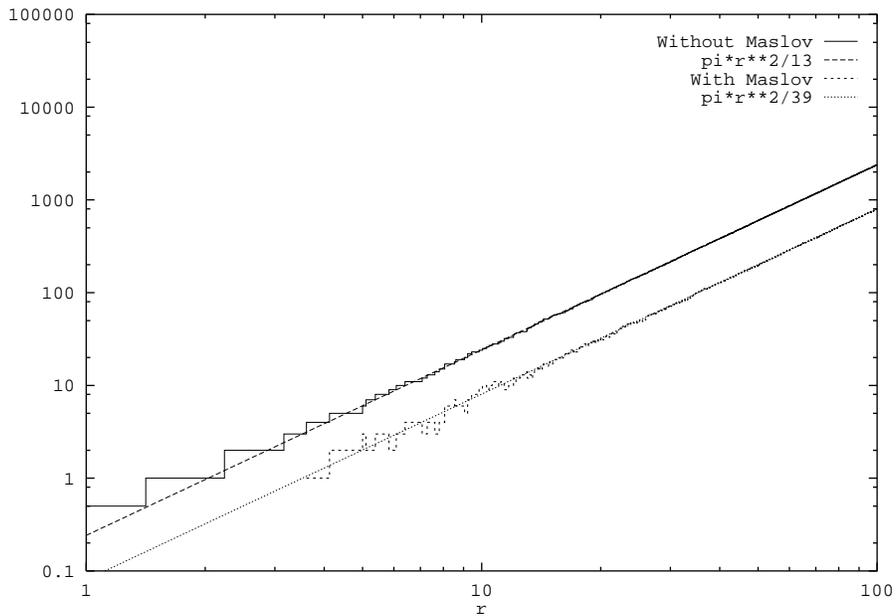}
\caption{The number of coprime lattice point, with and without {\em Maslov
weigths}, inside radius $r$, compared with the asymptotic results obtained in
the Appendix.}
\end{figure}

\subsection{Relation to cycle expansions}

All values of the symbol $s=(q,g)$ (introduced in section 3.2 and
corresponding to one laminar segment) can be realized if $q$ lies within the
horizon. An idea is now to build strings of these symbols and
build up arbitrary trajectories $s_1 s_2 s_3 \ldots$ and so define a symbolic
dynamics. If all the $s$'s lies inside the horizon almost all such sequences
could be realized and the symbolic dynamics would be almost complete.
This would yield a fundamental part of a cycle expansion
\begin{equation}
Z(k)=1-\sum_{s}
\frac{1}{|\Lambda_{\overline{s}}|^{1-\tau }}
e^{-ikl_{\overline{s}}} \ \ .
 \label{eqn:Zfund}
\end{equation}
If we compare this expression with eq \EqRef{Zhat} together with
eq \EqRef{aqfull} we see an apparent similarity. Indeed the prefactors are
almost the same, see \cite{PDSin}. (To make this comparison one first writes
the sum $\sum_{s=(q,g)}=\sum_q \sum_g$ and sums over $g$.)

The cycle expansion breaks down when some of the $s$'s are outside
the horizon whereas the BER approximation works fine.
The conclusion is that the approximation worked out in sections 3.1 to 3.3
provides a good generalization for the fundamental part of a cycle expansion.
Indeed it is both better (it automatically preserves unitarity) and simpler
as it does not use
an enourmous amount of periodic orbits to
explore the available phase space for one single symbol but rather measure it
directly. Neither does does it need all these cycles in order to find out that
the invariant density is uniform as we already knew that in advance.

\section{The Sinai billiard - Semiclassical considerations}

For the quantum Sinai billiard a major part is played by the (one parameter
families) of neutral periodic orbits.
The level density can be written as a sum
of the mean density and two oscillating parts, one from the unstable orbits
(expressed by the Gutzwiller formula) and one due to neutral orbits.
\begin{equation}
d(E) \equiv  \sum_{\nu} \delta (E-E_\nu) =
\bar{d}(E)+d_u^{osc}(E)+d_n^{osc}(E).
\end{equation}
The semiclassical approximation can be refined by taking
creeping orbits \cite{Creep} and ghost orbits \cite{Ghost}
into account , but we neglect them here.
We we will now discuss the two oscillating contributions separately.
In the discussion of the contribution from unstable orbits we will focuse on
convergence and singularities of the zeta function. The considerations
for neutral orbits are worked out for use in section 5.

\subsection{The contribution $d_u^{osc}(E)$ from unstable orbits}

We use the BER approximation to approximate the Gutzwiller-Voros zeta function
and we will use the rough approximations of section 3.3,
putting $\tau =1/2$,
and $\kappa=\sqrt{2E}=-k$.
According to our previous investigations we now expect this zeta function to
approximate the Gutzwiller-Voros zeta function for small
$k$, well below the first quantum state. This is certainly
not the semiclassical limit but will lead to some interesting observations.

We now have
to introduce
Maslov indices. The generalization of $p_{\tau}$ now includes an oscillating
weight
(we still restrict
us to the $A_1$ subspace):
\begin{equation}
\tilde{p}_{1/2}(\Delta) \approx \sum_{q=(n_x,n_y)}
(-1)^{n_x+n_y+1}a_q(\tau) \delta (\Delta-l_q) \ \ .
\end{equation}
The $a_q(\tau)$'s are the same as before but the definition of the
generalized density of
coprimes $\tilde{d}_c(r)$ must now include
the oscillating weight. The result appears to be exactly one third
of the previous result $\tilde{d}_c \sim \frac{2 \pi}{39}r^2$, see Appendix
and fig 6. The crude approximation of the probability distribution is
obtained by simply multiplying eq. \EqRef{pcrude} by one third.

The zeta function is obtained by Fourier transforming
our approximation of $\tilde{p}_{1/2}(\Delta)$.
An interesting question is if its leading zero is still below the real $k$
axis. In this approximation it follows from \EqRef{pcrude} (with the extra
factor of one third) that this is the case if
\begin{equation}
R<\frac{44 \; \Gamma (3/4)^2}{135 \; \sqrt{2} \; \Gamma(3/2)}\approx 0.3905
\end{equation}
For radius as big as the limiting value the approximation has already ceased
to be relevant. We must expect that the Euler product \EqRef{ZGV} is divergent
on the real axis and in a strip below, but the leading zero can be very close
to the real axis. It may very well happen that the zero crosses the real axis
for other subspaces.

If there is a zero below the real axis it will mean that the Gutzwiller trace
formula diverges on the real axis. In that case, whenever we encounter as
periodic orbit sum like \EqRef{tracedef} it must be regularized. This amounts
to keep the contour of the Fourier integral along the real axis. As all our
approximate zeta functions converges in the entire lower halfplane
this is not a major problem.

It is interesting to compare this situation
with the Selberg zeta function for systems with constant negative curvature.
Along the real $k(=-\kappa)$ axis the Selberg zeta function
has zeros on the exact (quantum)
locations, which is a peculiarity of these systems.
In addition it has a leading zero on the negative imaginary axis,
in exact analogy with our crude
estimations of the zeta functions for the Sinai billiard.
The situation is similar for the Riemann zeta function.

Our semiclassical zeta function has a cut along the positive imaginary axis
(and an infinite number of parallell cuts, cf. ref. \cite{PDSin}).
Do we allow the semiclassical zeta function to
have non analytical features? Well, the semiclassical approximation to the
spectral determinant $D(\kappa)$ is
$D(\kappa)\approx D_{sc}(\kappa ) =\exp (-i\pi
(\bar{N}+N_n^{osc}))Z_{GV}(\kappa)$.  Its exact
quantum counterpart $D(\kappa)$ obeys the functional equation
$D(\kappa)=D(-\kappa )$. There is nothing saying a priori that
$D_{sc}(\kappa)$ must obey this relation (it would be violated by
a cut), it could be an artifact reflecting the
shortcomings of the stationary phase approximation and/or the singularity of
the semiclassical limit.

But it should be noted that the best
semiclassical calculations for chaotic systems so far is obtained by imposing
the functional equation to the Gutzwiller-Voros zeta function.
It is shown \cite{func} that this requires a remarkable bootstrapping
property between short and long periodic orbits.
That such a bootstrapping indeed occurs is supported by the results of ref
\cite{SS}.
There are ways of
saving the semiclassical determinant from violating the functional equation.
The contribution from neutral orbits has cuts at the same places as
$Z_{GV}$ but they do not cancel, but the mean staircase
$\bar{N}(\kappa)$ could have  nonanalyticities cancelling the (main) cut of
$Z_{GV}$ and $N_n^{osc}$. This would extend the bootstrapping ideas
mentioned above as
$\bar{N}$ is governed by the short time behaviour and the cuts of $Z_{GV}$
by the long periodic orbits. Generally $\bar{N}$ is given by an asymptotic
expansion so it shouldn't be surprising if this expansion lead
to a cut.

In the classical case the cut carry information about the power law decay of
correlations \cite{PDSin}. It would be nice to know what it tells us in
the quantum case.


\subsection{The contribution $d_n^{osc}(E)$ from neutral orbits}

There is a neutral orbit in any free direction. The general formula is
\cite{Creagh}
\begin{equation}
d_n^{osc}=\sum^{+} 2B_i cos(\kappa l_i)   \label{eqn:dnosc}
\end{equation}
with
\begin{equation}
B_i=\frac{1}{(2\pi)^{3/2}} \sqrt{\frac{l_p}{|n_i|\kappa }}D_{p_i}
\equiv \tilde{B}_i  \frac{1}{\sqrt{\kappa}}   \label{eqn:Creagh}
\end{equation}
The index $i$ runs over all neutral periodic orbits.
Their length is written as
$l_i=n_i l_{p_i}$ where $n_i$ is a repetition number and $l_{p_i}$ the length
of the corresponding primitive periodic orbit.  $D_{p_i}$ is the geometrical
width of the orbit.
Note that this contribution is $O(1/E^{1/4})$ whereas the one due to
unstable orbtits is $O(1/E^{1/2})$. The reason that the unstable orbits
still are important is because they are so numerous. Generally the neutral
orbits affect the low part of the spectrum and large scale structures
in the higher part of the
spectrum.

For later puposes
it is convenient to sum over both positive and negative traversals
\begin{equation}
d_n^{osc}=\sum^{\pm} B_i e^{ i\kappa n_i l_{p_i}}   \label{eqn:dnpm}
\end{equation}
Instead of applying these equations directly
it is instructive to first consider the integrable $R
\rightarrow 0$ limit of the Sinai billiard. The wavefunction must now vanish
at the midpoint which is the relic of the disk.
The spectrum is given by
$E=2\pi^2(n^2+m^2)$ where $m,n$ are postive integers. This leads to a mean
level density $\bar{d}=1/8\pi$ which differs form the Weyl term
$\bar{d}=1/2\pi$. The reason is that is that the $R \rightarrow 0$ is a very
odd system quantum mechanically, only having wave functions in two of the
symmetry classes ($A_2$ and $B_2$).

To find $d_n^{osc}(E)$ we perform the
Berry-Tabor trick \cite{BClust} and rewrite the level density
by means of the
Poisson's summation formula
\begin{equation}
d_n(E)=\sum_{m,n} \delta (E-2\pi^2(m^2+n^2))=
\sum_{M,N=-\infty}^{\infty} \int dn \;dm \; \delta (E-2\pi^2(m^2+n^2))
e^{2\pi i (Nn+Mm)}
\end{equation}
Performing the integral by stationary phase we arrive at
\EqRef{dnosc} with
\begin{equation}
\tilde{B}_i=\frac{1}{(2\pi )^{3/2} \sqrt{N^2+M^2}}
\end{equation}
and
\begin{equation}
l_i=\sqrt{N^2+M^2}
\end{equation}
where $i$ ranges over all pairs $(M,N)$ in the first quadrant.
Only one symmetry class (say $A_2$) is considered if we restrict the
summation to the first octant
(not bothering how to treat lattice points on the symmetry line $M=N$).
If we now write
the lattice vector as $(M,N)=n(n_x,n_y)$ where
$(n_x,n_y)$ are coprime, we will get consistency with \EqRef{Creagh} if
$l_p=\sqrt{n_x^2+n_y^2}$ and $D_p=1/l_p$.

The width $D_p$ decreases as the radius $R$ increases. Inspecting
\EqRef{corrr} we realize that the width then becomes
$D_p=1/l_p-2R$. These results are derived in ref. \cite{BSin} in another way.

With the equipment procured so far we will now attempt to
calculate  the spectral form factor.

\section{More semiclassics - The spectral form factor}

A lot of information about spectral fluctuations is encoded in the spectral
form factor \cite{BRig}
\begin{equation}
K({\cal T})=\frac{1}{\bar{d}} \int_{-\infty}^{\infty}
d\epsilon e^{i2\pi \bar{d} {\cal T} \epsilon}
\langle d^{osc}(E-\frac{\epsilon}{2})d^{osc}(E+\frac{\epsilon}{2})
\rangle_E  \ \ .
\label{eqn:Kdef}
\end{equation}
The averaging $\langle \ldots \rangle_E$ has been performed in a
small range $\Delta E$ such that $\bar{d}^{-1} \ll \Delta E \ll E$.
If there are contributions from both neutral and unstable orbits
the form factor divides into three parts (with obvious notations)
\begin{equation}
K({\cal T})=K_{nn}({\cal T})+K_{uu}({\cal T})+K_{un}({\cal T})
\end{equation}

The suggested universal form factor relevant for time reversable chaotic
systems is that of the Gaussian Orthogonal Ensamble (GOE) which reads
\cite{Meht}
\begin{equation}
K_{GOE}= \left\{ \begin{array}{ll}
2{\cal T}-\tau \log(1+2{\cal T}) & {\cal T} \leq 1 \\
2-{\cal T} \log \frac{2{\cal T} +1}{2{\cal T} -1}  &
{\cal T} \geq 1 \end{array}
\right.
\end{equation}

\begin{figure}
\epsffile{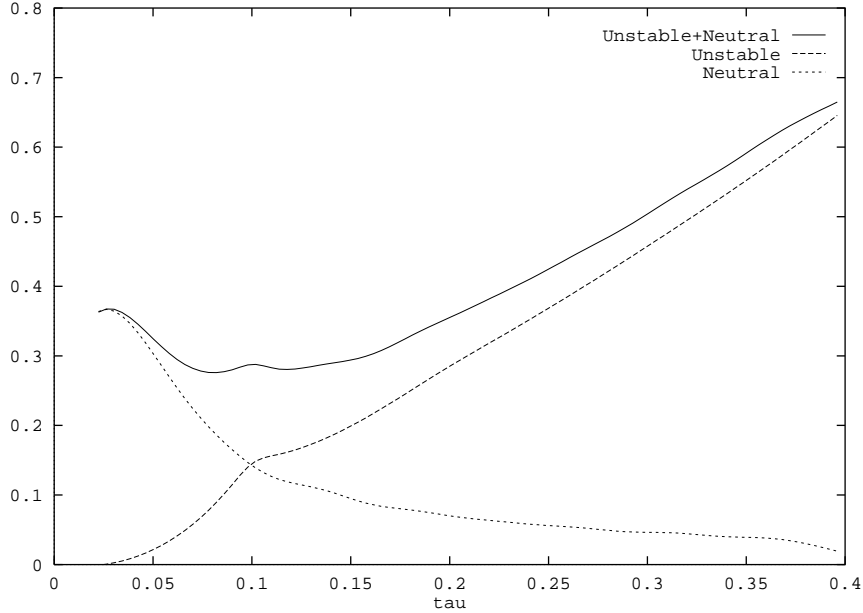}
\caption{The contributions $K_{nn}$ and $K_{uu}$ and their sum.
The latter is calculated from eq (44) using the methods of section 3.3.
The parameters are $R=0.1$ and ${\cal T}_{scale}=0.1$.}
\end{figure}

\subsection{The contribution from neutral orbits: $K_{nn}({\cal T})$}

If we insert $d_n^{osc}$
into the definition of $K_{nn}({\cal T})$, and expanding
$\sqrt{2E+\epsilon}\approx \sqrt{2E}+\epsilon/\sqrt{2E}$ we obtain
\begin{equation}
K_{nn}({\cal T} )=
\frac{2\pi}{\bar{d}}  \sum_i^{\pm} \sum_j^{+} \tilde{B}_i
\tilde{B}_j  \delta (l-\frac{l_i+l_j}{2})
\langle cos(\sqrt{2E}(l_i-l_j)) \rangle_E
   \ \ .
\label{eqn:Knn}
\end{equation}
where the length variable $l({\cal T})=\sqrt{2E}2\pi \bar{d} {\cal T}$.
This expression is always dominated by the diagonal terms \cite{BClust}
\begin{equation}
K_{nn}({\cal T} )\approx
\frac{2\pi}{\bar{d}} \sum_i \tilde{B}_i^2 \delta(l-l_i) \ \ .
\end{equation}
For an integrable system this always equals unity \cite{BClust}.
This can be easily checked for
the Sinai $R \rightarrow 0$ billiard by replacing the sum $\sum_{M,N}$ by an
integral and using polar coordinates.

If $R$ is very small and $l$ is far from
$l_{scale}$, $K_{nn}$ will still be close to unity, but it will now decrease
with increasing ${\cal T}$. An example is given in fig. 7.

Let us now study the case $l \gg l_{scale}$. We can be more general and
consider and arbitrary system with a finite number of primitive neutral
orbits in the limit $l \gg \max_p l_p$( $= l_{scale}$ for
the Sinai case) we have
\begin{equation}
K_{nn}({\cal T})=\frac{1}{(2\pi)^2 \bar{d}} \sum_p l_p D_p^2 \sum_n
\frac{1}{n}
 \delta(l-nl_p)=
\frac{1}{(2\pi)^2\bar{d}}\frac{1}{l({\cal T})} \sum_p l_p D_p^2 \ \ .
\end{equation}
We thus have an inverse power law decay,
$K_{nn} \sim \frac{1}{\sqrt{E}{\cal T}}$
in contradiction to the large ${\cal T}$ limit of the
GOE prediction $K_{GOE} \sim 1-\frac{1}{12{\cal T}^2}$.

\subsection{The contribution from unstable orbits: $K_{uu}({\cal T})$}

In the limit ${\cal T} \ll 1$ this contribution to the
form factor is also dominated
by the  diagonal terms. However, the result must be multiplied by
two \cite{BRig} since
most periodic orbits in time reversable systems
are rotations, and each rotation counts as two cycles with exactly the same
length.  We insert the trace formula into the
definition of the form factor
and express the result in terms of the trace of the evolution operator
\begin{equation}
K_{uu}({\cal T}) = 2 {\cal T} tr {\cal L}^{l({\cal T})}_{\tau=0}
\end{equation}
An example of this function is found in fig 7.
If ${\cal T}_{scale} \ll {\cal T} \ll 1$ we find, after consulting
eqs. \EqRef{pcrude} and \EqRef{tras}, that
$K_{uu}=2{\cal T}-2{\cal T}_{scale}$ where the first term agrees
with the universal GOE result \cite{BRig} and the second being
a correction dying
out in the limit $E \rightarrow \infty$.
${\cal T}_{scale}$ is the scale
corresponding to $l_{scale}$ given by
 ${\cal T}_{scale}=l_{scale}/(\sqrt{2E}2\pi \bar{d})$.

The big challenge for semiclassical methods is to calculate the large
${\cal T}$ limit of $K_{uu}({\cal T})$. General considerations
\cite{BRig} say that the form factor must approach unity as
${\cal T} \rightarrow \infty$.
The statement is often written in terms of an
explicit periodic orbit double sum analogous to \EqRef{Knn}.
One should always be extremely
careful when encountering such sums since they in general require
regularization {\em even if the double sum is convergent}; otherways one will
obtain an exponentially increasing form factor \cite{AS}.

\begin{figure}
\epsffile{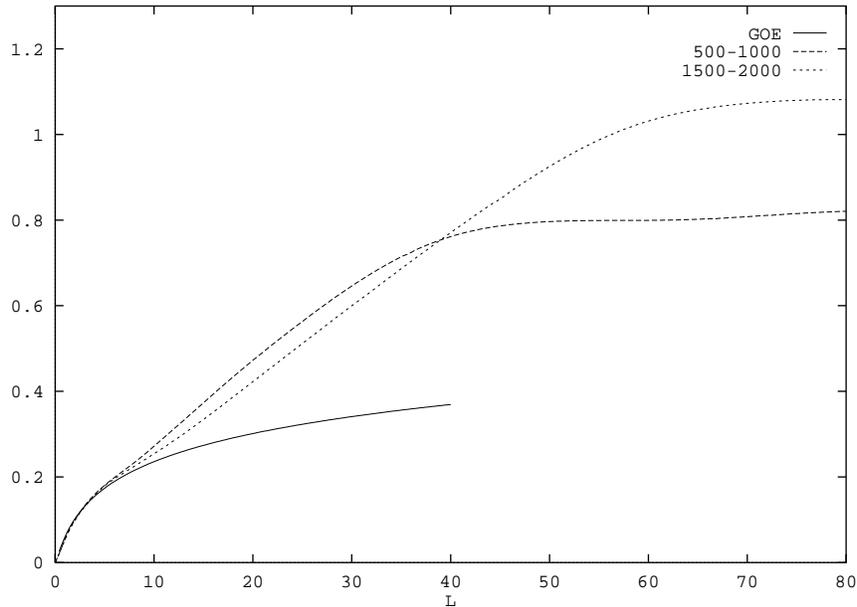}
\caption{The spectral rigidity of the Sinai billiard calculated from two
subsets of the spectrum and compared with GOE. The radius $R=0.1$.
The relevant scale ${\cal T}_{scale}$ is slightly above (below)
${\cal T}_{scale}=0.1$ for
the lower (higher) part of the spectrum.
}
\end{figure}

\vspace{0.5cm}

We will not calculate the mixed contribution $K_{un}$; it may provide small
corrections. It seems unlikely that $K_{un}$ and/or $K_{uu}$ could cancel
the $1/{\cal T}$ decay of $K_{nn}$.
Summing up our contributions, fig 7, we conclude that
for any finite energy the
form factor is not compatible with GOE
for small ${\cal T}$ either.
The
small ${\cal T}$ limit governs the large scale structure of the spectrum.
A convenient mesaure to study such properties is the spectral rigidity which
we now will discuss.

\subsection{Spectral rigidity}

Spectral rigidity, or $\Delta_3(L)$ statistics \cite{Dys},
measures the (mean square)
deviation of the spectral staircase from a linear function over $L$ mean
spacings.
Small values of $\Delta_3(L)$ thus correspond to a
rigid spectrum. $\Delta_3(L)$ is expressed in terms of the form factor
as \cite{BRig}
\begin{equation}
\Delta_3 (L)=\frac{1}{2\pi^2}
\int_{0}^{\infty}\frac{d{\cal T}}{{\cal T}^2}K({\cal T})G(\pi L {\cal T})
\end{equation}
where
\begin{equation}
G(y)=1-\frac{1}{y^2}-2\frac{cos^{2}(y)}{y^2}+6\frac{cos(y)sin(y)}{y^3}
-3\frac{sin^{2}(y)}{y^4}
\end{equation}
The function $G(y)$ is close to unity if $y >1$ so for sufficiently
big $L$ the rigidity explores the small ${\cal T}$ limit of the form factor.

The deviation from GOE for small ${\cal T}$ implies that even {\em if}
$\Delta_3(L)$ follows GOE for a while it must certainly deviate from it
well before $L=1/{\cal T}_{scale}=2R\sqrt{2E}2\pi\bar{d}$.
To study this we have also computed the exact quantum spectrum by the methods
of ref. \cite{BSin}. It is clearly seen in fig. 8 that such
deviation occurs well before the saturation value $L_0$ corresponding to the
shortest periodic orbit $l_0$. Such  behavior has been observed before
\cite{Arve}. If the form factor approaches a constant $c$ as ${\cal T}
\rightarrow 0$, then the rigidity will be asymptotically linearly increasing
with slope $c/15$. The slope in fig. 8 is roughly $0.17$ corresponding to a
constant $c \approx 0.25$, which is quite compatible with the results of fig 7.
As we said, we expect $c$ to approach unity when the disk radius
$R \rightarrow 0$. Accordingly the slope of $\Delta_3$ should approach
$1/15$ (as for an integrable system). That this is indeed the case is
supported by the results of ref. \cite{Arve}.

The effect we have seen is mostly due to the neutral orbits.
However, we can expect similar deviations, but perhaps less pronounced,
for systems without neutral orbits. The reason is that the asymptotics of the
trace (which was a slow power law for the Sinai billiard) does not set in
until a length scale $l_{scale}$ which can be much longer than the shortest
periodic orbit $l_0$. For instance, in the hyperbola billiard $l_0=1$
whereas $l_{scale} \approx 27.61$, see ref. \cite{PDreson}.

\section{Concluding remarks}

We have surely not used the full potential of the methods outlined in section
3. I conclude by listing some lines along which the work
may proceed.

\begin{itemize}

\item Improve the BER approximation by replacing the delta functions in sec
3.1 by something more realistic.

\item Find corrections (similar to curvature terms) to the BER approximation.
One should also correct for the nonmultiplicativity of the weight.

\item Try to calculate the long ${\cal T}$ limit of the form factor. It is
thus essential to have good control of the fine scale structure
the (properly regularized) trace $tr {\cal L}_{\tau=1/2}^l$. The requires
that the previous two points have been worked out. Is there perhaps any simple
universal structure in this fine structure?

\item Work out the BER approximation (with corrections) for other systems.
Each system will of course present its own problems: the stadium billiard has
no natural division between a chaotic and a regular part, as it is glued
together from regular components. The anistropic Kepler problem \cite{Gregor}
and the closed
three disk billiard suffers from infinite sequences of cycles accumulating to
finite length and stability. The hyperbola billiard has infinite horns.

\end{itemize}

This work was supported by the Swedish Natural Science
Research Council (NFR) under contract no. F-FU 06420-303.
I will also thank Predrag Cvitanovi\'{c} for various remarks.

\section*{Appendix}

The number of coprime lattice points inside radius $R$ in the first
octant, $N_c(R)$
\begin{equation}
N_c(r)=\sum_{gcd(n_x,n_y)=1}
\Theta(r-\sqrt{n_x^2+n_y^2})
\end{equation}
is given by the integral equation
\begin{equation}
\frac{\pi R^2}{8}-N_c(R)=
\int_0^{R/2} \frac{dN_c(r)}{dr}([\frac{R}{r}]-1)dr
\end{equation}
This equation expresses the fact that density of coprimes at radius
r contributes to the density of non-coprimes ar radii $2r$, $3r$ etc.
Each non-coprime is uniquely related to one coprime point. This summability
leads to a simple integral as above.
The total number of lattice points inside radius $R$ is $\pi R^2/8$,
appearing in the left hand side. $[x]$ denotes the integral part of $x$, which
is on the average $=x-1/2$. Inserting this into the equation above it is easily
seen that $N_c(r)= \frac{\pi}{13}r^2$ is a solution.

The introduction of maslov indices in section 4.1 motivates us to study
the asymptotic behaviour of
\begin{equation}
\tilde{N}_c(r)=\sum_{gcd(n_x,n_y)=1} (-1)^{n_x+n_y+1}
\Theta(r-\sqrt{n_x^2+n_y^2})
\end{equation}
Numerical calculation suggests that $\tilde{N}_c(r) \sim 1/3 \cdot N_c(r)$.
How can this be understood? The pairs $(n_x,n_y)$ can either be
(odd,odd), (odd, even), (even,odd). The (even,even) option is
excluded by not being coprime.
If the three possibilities have the same density the
result follows.

\newcommand{\PR}[1]{{Phys.\ Rep.}\/ {\bf #1}}
\newcommand{\PRL}[1]{{Phys.\ Rev.\ Lett.}\/ {\bf #1}}
\newcommand{\PRA}[1]{{Phys.\ Rev.\ A}\/ {\bf #1}}
\newcommand{\PRD}[1]{{Phys.\ Rev.\ D}\/ {\bf #1}}
\newcommand{\PRE}[1]{{Phys.\ Rev.\ E}\/ {\bf #1}}
\newcommand{\JPA}[1]{{J.\ Phys.\ A}\/ {\bf #1}}
\newcommand{\JPB}[1]{{J.\ Phys.\ B}\/ {\bf #1}}
\newcommand{\JCP}[1]{{J.\ Chem.\ Phys.}\/ {\bf #1}}
\newcommand{\JPC}[1]{{J.\ Phys.\ Chem.}\/ {\bf #1}}
\newcommand{\JMP}[1]{{J.\ Math.\ Phys.}\/ {\bf #1}}
\newcommand{\JSP}[1]{{J.\ Stat..\ Phys.}\/ {\bf #1}}
\newcommand{\AP}[1]{{Ann.\ Phys.}\/ {\bf #1}}
\newcommand{\PLB}[1]{{Phys.\ Lett.\ B}\/ {\bf #1}}
\newcommand{\PLA}[1]{{Phys.\ Lett.\ A}\/ {\bf #1}}
\newcommand{\PD}[1]{{Physica D}\/ {\bf #1}}
\newcommand{\NPB}[1]{{Nucl.\ Phys.\ B}\/ {\bf #1}}
\newcommand{\INCB}[1]{{Il Nuov.\ Cim.\ B}\/ {\bf #1}}
\newcommand{\JETP}[1]{{Sov.\ Phys.\ JETP}\/ {\bf #1}}
\newcommand{\JETPL}[1]{{JETP Lett.\ }\/ {\bf #1}}
\newcommand{\RMS}[1]{{Russ.\ Math.\ Surv.}\/ {\bf #1}}
\newcommand{\USSR}[1]{{Math.\ USSR.\ Sb.}\/ {\bf #1}}
\newcommand{\PST}[1]{{Phys.\ Scripta T}\/ {\bf #1}}
\newcommand{\CM}[1]{{Cont.\ Math.}\/ {\bf #1}}
\newcommand{\JMPA}[1]{{J.\ Math.\ Pure Appl.}\/ {\bf #1}}
\newcommand{\CMP}[1]{{Comm.\ Math.\ Phys.}\/ {\bf #1}}
\newcommand{\PRS}[1]{{Proc.\ R.\ Soc. Lond.\ A}\/ {\bf #1}}

\end{document}